\documentclass[aps,prl,floatfix, twocolumn, superscriptaddress]{revtex4}
\usepackage[makeroom]{cancel}
\usepackage{bm}
\usepackage{amsmath}
\usepackage{mathtools}
\usepackage{amsfonts}
\usepackage{siunitx}
\usepackage{graphicx}

\newcommand{\be}{\begin{equation}}
\newcommand{\ee}{\end{equation}}

\usepackage{hyperref}
\usepackage{adjustbox} 
\usepackage[dvipsnames]{xcolor}
\begin{document}

\title{Magnetization dynamics in a three-dimensional interconnected nanowire array}
\author{Rajgowrav Cheenikundil}
\affiliation{Universit{\'e} de Strasbourg, CNRS, Institut de Physique et Chimie des Mat{\'e}riaux de Strasbourg, F-67000 Strasbourg, France}
\email{riccardo.hertel@ipcms.unistra.fr}
\author{Massimiliano d'Aquino}
\affiliation{Department of Electrical Engineering and ICT, University of Naples Federico II, Naples, Italy}
\author{Riccardo Hertel}
\affiliation{Universit{\'e} de Strasbourg, CNRS, Institut de Physique et Chimie des Mat{\'e}riaux de Strasbourg, F-67000 Strasbourg, France}
\date{\today}
\begin{abstract}
Three-dimensional magnetic nanostructures have recently emerged as artificial magnetic material types with unique properties bearing potential for applications, including magnonic devices. Interconnected magnetic nanowires are a sub-category within this class of materials that is attracting particular interest. We investigate the high-frequency magnetization dynamics in a cubic array of cylindrical magnetic nanowires through micromagnetic simulations based on a frequency-domain formulation of the linearized Landau-Lifshitz-Gilbert equation. The small-angle high-frequency magnetization dynamics excited by an external oscillatory field displays clear resonances at distinct frequencies. These resonances are identified as oscillations connected to specific geometric features and micromagnetic configurations. The geometry- and configuration-dependence of the nanowire array's absorption spectrum demonstrates the potential of such magnetic systems for tuneable and reprogrammable magnonic applications. 
\end{abstract}
\maketitle

\section{Introduction}
Over many years, cylindrical soft-magnetic nanowires have been the subject of intense research in micro- and nanomagnetism \cite{vazquez_magnetic_2015}. Their magnetic properties have been analyzed regarding the structure and motion of domain walls \cite{forster_domain_2002,hertel_magnetization_2004, yan_beating_2010, wieser_domain_2004}, which, depending on the nanowire thickness, can display remarkable stability and high propagation velocities \cite{yan_fast_2011,hertel_ultrafast_2016}. Magnetic nanowires and nanotubes have also attracted interest within the magnonics community \cite{wang_sub-50_2021}
and in the nascent field of curvilinear micromagnetism \cite{makarov_curvilinear_2022}, as it was found that, due to their rounded cylindrical outer shape, these geometries can exhibit non-reciprocal spin-wave propagation \cite{otalora_curvature-induced_2016}. In addition to studies on individual nanowires, research has also been conducted on the collective behavior of magnetostatically interacting nanowire arrays \cite{hertel_micromagnetic_2001, clime_magnetostatic_2006}. Recently, structurally even more complex types of ensembles have been investigated in the form of arrays of {\em interconnected} nanowires \cite{burks_3d_2021}. In these systems, the intersection sites of crossing wires introduce additional geometric and micromagnetic features, which provide further potential for functionalities in future magnetic nanodevices \cite{frotanpour_vertex_2021, saavedra_dynamic_2020, cheenikundil_high-frequency_2022}. Such interconnected nanowires arrays can be fabricated, e.g., by oblique ion irradiation of a polymer matrix followed by a filling of the generated pores with magnetic materials through electrodeposition \cite{da_camara_santa_clara_gomes_making_2019}, which results in extended, irregular networks with statistically distributed spacing between nanowires. Alternatively, by means of layer-by-layer nanofabrication techniques such as FEBID (focussed electron-beam induced deposition), it is possible to generate smaller but highly regular arrays of connected magnetic nanowires \cite{teresa_review_2016, keller_direct-write_2018, skoric_layer-by-layer_2020}.

Extending the research from studies on single nanowires to interconnected arrays not only expands the knowledge of the magnetic properties of nanowires but also contributes to furthering the more general topic of three-dimensional (3D) nanomagnetism---an emerging field of research \cite{fernandez-pacheco_three-dimensional_2017, fischer_launching_2020} which
aims at identifying the impact of nanoscale 3D geometric features on 
the magnetization.
The ability to fabricate arbitrarily shaped nanomagnets and nanoarchitectures \cite{fernandez-pacheco_writing_2020, keller_direct-write_2018, gliga_architectural_2019} may
open up a plethora of possibilities, which could ultimately enable magnetic materials with tailored properties governed by their geometric features. 
Potential applications of 3D nanomagnetic materials are foreseen, e.g., in the field of magnonics \cite{gubbiotti_three-dimensional_2019}, high-density data storage, and neuromorphic computing \cite{grollier_neuromorphic_2020}.

\begin{figure}[htpb]
\includegraphics[width=\linewidth]{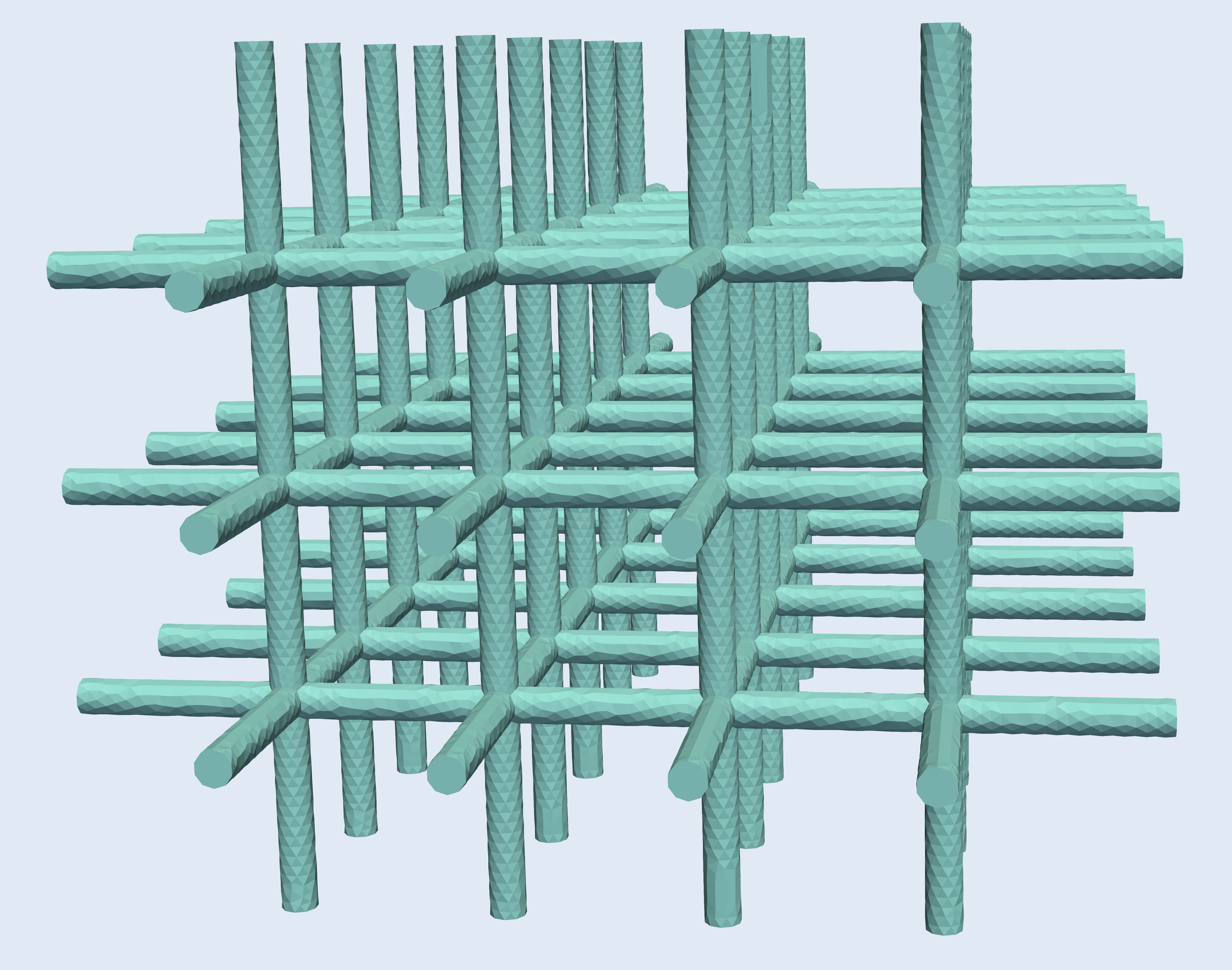}
\caption{\label{fig:sample} Perspective view on the studied sample geometry. The array consists of three orthogonal sets of equidistant cylindrical magnetic nanowires. The intersection points constitute a regular cubic lattice with a constant of \SI{70}{\nano\meter}. The structure's total size is \SI{420}{\nano\meter}$\times$\SI{350}{\nano\meter}$\times$\SI{280}{\nano\meter}.}  
\end{figure}

Within this type of artificial magnetic material, regular arrays of interconnected nanowires represent an interesting subcategory athat combines the unique micromagnetic features of cylindrical nanowires with the magnonic properties of 3D nanoarchitectures. In addition, since the dominant magnetostatic effect tends to orient the magnetization along the nanowire axis, the arrays are akin to assemblies of Ising-type nanomagnets, which imparts to these systems the characteristics of 3D artificial spin ices \cite{sahoo_observation_2021}. 
These qualities render arrays of magnetic nanowires 
a complex class of systems whose study could pave the way toward the design of three-dimensional magnetic metamaterials \cite{hertel_defect-sensitive_2022}.
Understanding the high-frequency oscillatory magnetization dynamics and its dependence on the magnetic structure and the geometry could help identify the potential of such material in the field of three-dimensional magnonics \cite{gubbiotti_three-dimensional_2019}.
\section{Model system}
As an example of this type of artificial magnetic material, we consider here a regular array of interconnected nanowires, as illustrated in Fig.~\ref{fig:sample}.
We investigate, through advanced micromagnetic simulations, the frequency-dependent small-angle oscillatory magnetization dynamics excited by a low-amplitude time-harmonic external field, and how this dynamic response depends on the static magnetic configuration of the nanowire array. Our sample consists of 47 magnetic nanowires, arranged in a regular $5\times 4\times 3$ grid. Twelve of them are aligned along the $x$ direction, 15 are parallel to the $y$ axis, and 20 are oriented in the $z$ direction. The intersection sites are located on a regular cubic grid, with a lattice constant of \SI{70}{\nano\meter}. Each cylindrical nanowire has a diameter of \SI{12}{\nano\meter} and their length varies between \SI{280}{\nano\meter} and \SI{420}{\nano\meter}, depending on the wire orientation. The geometric parameters of this array are compatible with what is currently achievable using modern FEBID nanofabrication technologies. Accordingly, we use typical material parameters of FEBID-deposited amorphous cobalt, i.e., an exchange constant of $A=\SI{15}{\pico\joule\per\meter}$, a spontaneous magnetization of $\mu_0M_s=\SI{1.2}{\tesla}$, and zero magnetocrystalline anisotropy \cite{AmalioCommunication} ($\mu_0$ is the vacuum permeability).

As a  model for the simulation, we use an unstructured tetrahedral finite-element mesh consisting of more than 170\,000 cells, and approximately 57\,000 nodes. The edge lengths of the tetrahedra are smaller than \SI{4}{\nano\meter}, which is below the material's exchange length $l_s=\sqrt{2A/\mu_0 M_s^2}\simeq\SI{5.1}{\nano\meter}$ and yields sufficiently small elements to obtain a good geometric approximation of the rounded nanowire shape.

\section{Simulation method\label{sec:method}}
To analyze the static and the oscillatory high-frequency micromagnetic properties of the system, we employ a two-step approach in which the two aspects are treated separately. In the first step, we use our general-purpose finite-element micromagnetic simulation software {\tt tetmag} to determine the zero-field equilibrium structure of the magnetization. The equilibrium state at zero field is not unique and depends on the sample's magnetic history. Analogous to the typical situation in artificial spin-ice lattices, a large number of quasi-degenerate configurations are possible, which can be characterized by the magnetic structures developing at the vertices, as will be discussed in section \ref{vertex_configs}. 

The static equilibrium is simulated by numerically integrating the Landau-Lifshitz-Gilbert (LLG) equation 
\cite{gilbert_phenomenological_2004}
\be \label{gilbert}
\frac{d\bm{M}}{dt}=-\frac{1}{1+\alpha^2}\left[\gamma\left(\bm{M}\times\bm{H}_\text{eff}\right)+\frac{\alpha}{M_s}\left[\bm{M}\times\left(\bm{M}\times\bm{H}_\text{eff}\right)\right]\right]
\ee
until convergence is reached, where $\alpha$ is the Gilbert damping constant, $\gamma$ is the absolute value of the gyromagnetic ratio, and $\bm{H}_\text{eff}$ is the micromagnetic effective field \cite{brown_micromagnetics_1963} obtained from the variational derivative of the micromagnetic free energy functional, which contains contributions from the ferromagnetic exchange, the magnetostatic (dipolar) interaction, and, if applicable, any static external field.
In practice, the time integration of the LLG equation is continued until the value of the local torque $\left|\bm{M}\times \bm{H}_\text{eff} \right|$ drops below a user-defined threshold at each discretization point. 
This first part of the calculation is a standard micromagnetic simulation task, which we perform reliably using numerical methods that we have developed over many years and used in numerous previous works.

In a second step, once the static equilibrium configuration is computed, we simulate the stationary and frequency-dependent response of the system to a weak externally applied sinusoidal field, $\delta\bm{h}_\text{ext}(t)=\delta\hat{\bm{h}}_\text{ext}\exp(i\omega t)$. As a convention, we refer to vector fields by lower-case letters when they are represented in reduced units, i.e., $\bm{h}_\text{ext}=\bm{H}_\text{ext}/M_s$, $\bm{m}=\bm{M}/M_s$, etc. Moreover, we use variables prefixed by a $\delta$ to denote oscillating, small-amplitude quantities, a zero subscript for static quantities, and the circumflex diacritic to denote complex oscillation amplitudes associated with small-amplitude fields. The applied time-harmonic field $\delta\bm{h}_\text{ext}$ excites small-angle oscillations of the magnetization around the equilibrium state $\bm{m}_0$, i.e.,
$\bm{m}=\bm{m}_0+\delta\bm{m}$, which are accompanied by fluctuations of the effective field $\bm{h}_\text{eff}=\bm{h}_0^\text{eff}+\delta\bm{h}_\text{eff}+\delta\bm{h}_\text{ext}$ (we explicitly separate the external applied field $\delta\bm{h}_\text{ext}$ from the remaining terms $\delta\bm{h}_\text{eff}$ in $\bm h_\text{eff}$). Inserting this perturbative approach 
into the LLG equation, assuming $\left|\delta\bm{m}\right|\ll\left|\bm{m}_0\right|$, results in  
\be\label{linear}
-i\omega\left(\alpha\delta\hat{\bm{m}}+\bm{m}_0\times\delta\hat{\bm{m}}\right)=
\mathcal{P}\cdot\left(h_0\delta\hat{\bm{m}}-\delta\hat{\bm{h}}_\text{eff}-\delta\hat{\bm{h}}_\text{ext}\right)
\ee
where $h_0=\bm{m}_0\cdot\bm{h}_0^\text{eff}$ and $\mathcal{P}={\cal I}-\bm{m}_0\otimes\bm{m}_0$. 


In discretized form, eq.~(\ref{linear}) yields a linear system that can be solved for $\delta\hat{\bm{m}}$ for any given frequency $\omega$ and external field amplitude $\delta\hat{\bm{h}}_\text{ext}$. In this linearized dynamics approach, the micromagnetic constraint of constant magnitude, $\left|\bm{m}\right|=1$, translates into the condition $\delta\bm{m}\cdot\bm{m}_0=0$. Therefore, 
the variations $\delta\bm{m}$ at each discretization point contain only two degrees of freedom (not three), as only variations perpendicular to $\bm{m}_0$ are admissible. Correspondingly, assuming that the numerical model contains $N$ discretization points, the system of equations to calculate $\hat{\bm{m}}$ needs to be solved for $2N$ unknowns. The reduction of variables from $3N$ to $2N$ is achieved by applying suitable coordinate transforms through local rotation operators \cite{daquino_novel_2009}.
Note that the numerical solution of the discrete counterpart of eq.~\eqref{linear} would normally involve a dense matrix with $\mathcal{O}(N^2)$ dimensions, which becomes soon unpractical even for a moderate number of computational cells. However, we achieve a large-scale matrix-free solution of eq.~\eqref{linear} by
using an operator-based formalism \cite{daquino_novel_2009, daquino_micromagnetic_2023},
which allows to preserve the almost linear $\mathcal{O}(N)$ complexity of the effective-field calculation \cite{hertel_large-scale_2019}. 

The solution of the linear system provides the frequency-dependent dynamic oscillation profile of the magnetization, 
\be 
\delta\bm{m}(\bm{x},t)=\delta\hat{\bm{m}}(\bm{x})\cdot\exp(i\omega t)\quad,
\ee 
which is obtained directly in the frequency domain, i.e., without performing a time-consuming integration of the LLG equation. To solve problems of this type, we have recently developed a powerful software package, the matrix-free micromagnetic linear-response solver (MF-$\mu$MLS) \cite{daquino_micromagnetic_2023}.

For this part of the simulation in the frequency domain, we assume a Gilbert damping of $\alpha=0.01$ and a weak field amplitude $\left|\delta\hat{\bm{h}}_\text{ext}\right|\mu_0 M_s=\SI{0.5}{\milli\tesla}$. In our simulations, the frequency $\omega$ is varied in steps of \SI{50}{\mega\hertz} in a range from \SIrange{1}{30}{\giga\hertz}, and the field amplitude $\delta\hat{\bm{h}}_\text{ext}$ is oriented along the $x$ axis.

\section{Static zero-field configurations\label{vertex_configs}}
The array of interconnected nanowires shown in Fig.~(\ref{fig:sample}) can be regarded as an assembly of 227 nanowire segments connected to the intersection points (the vertices). Each segment behaves like an Ising-type magnet, i.e., it can be magnetized in one of two possible axial directions. Accordingly, a large number of magnetization states can be attained---a situation that is well-known from artificial spin-ice systems \cite{nisoli_ground_2007, perrin_extensive_2016}. To reduce the complexity related to the numerous possible magnetic configurations, we consider only three qualitatively different states, the analysis of which will help us identify the general properties of the system. 

\begin{figure}[ht]
\includegraphics[width=\linewidth]{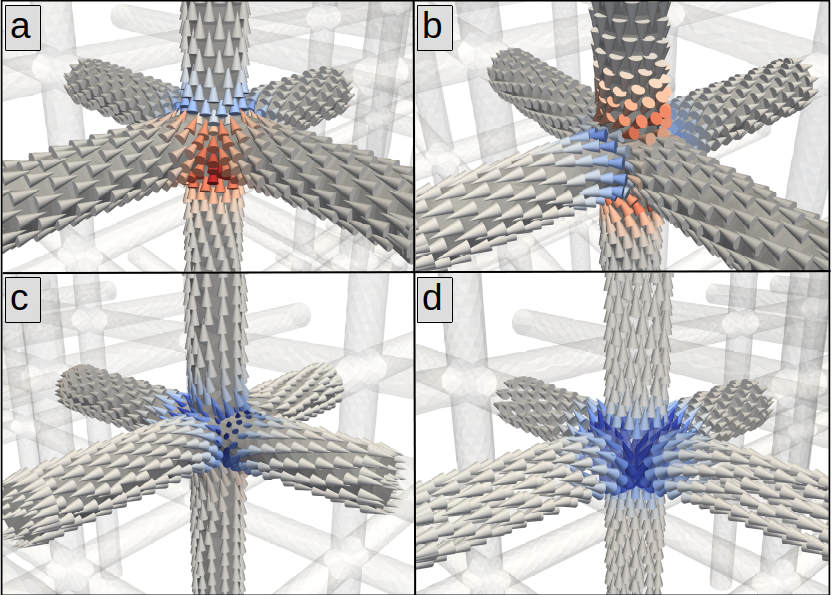}
\caption{\label{fig:vertices} Four qualitatively different vertex configurations: (a) uncharged state of type I, (b) uncharged state of type II, (c) double-charge configuration, and (d) quadruple-charge configuration. The color of the arrows displays the local magnetostatic volume charge density $\rho=-\nabla\cdot\bm{M}$, with red representing positive and blue negative values.}  
\end{figure}

The first configuration, which we call the reference state, is one in which the magnetization in each wire is uniform, such that all wire segments belonging to the three orthogonal sets are magnetized in the same way, i.e., along the positive $x$, $y$, and $z$ axis, respectively.
As a second configuration, which we call the charged state, we select a remanent state that develops at zero field after saturating the sample in the $x$ direction. In this configuration, all nanowire segments oriented parallel to the $x$ axis remain magnetized in the positive $x$ direction. The magnetization in the other two sets of wires, in contrast, is not necessarily uniform, and the corresponding nanowire segments can be either magnetized along the positive or negative $y$ and $z$ directions, respectively. Finally, as a third state, we consider a zero-field magnetic configuration that develops after numerically relaxing the system to equilibrium when starting from a fully randomized initial magnetic structure. In this random state, the magnetization is {\em a priori} different in each nanowire segment, pointing towards the negative or positive $x$, $y$, and $z$ directions. In practice, however, not all combinations are possible as certain local magnetic configurations are energetically unstable, in particular the ``hedgehog''-type structure where the magnetization in all six adjacent wires points towards or away from a vertex.

Different magnetic configurations can develop at the vertices throughout the array. As the magnetization in adjacent nanowire segments points towards or away from a vertex, it carries positive or negative magnetic flux toward the intersection site. An imbalance in the number of nanowire segments magnetized toward and away from the vertex generates an effective magnetic charge, which makes it possible to distinguish between charged and uncharged vertex sites. 
The charge can be visualized by plotting the divergence of the magnetization field, which, except for a sign convention, is equivalent to the definition of the magnetostatic volume charge density \cite{brown_magnetostatic_1962}. 

Figure \ref{fig:vertices} shows the four basic magnetic vertex configurations found in the array. Frames (a) and (b) display two different zero-charge structures ($0q$), with a ``three-in/three-out'' magnetic configuration in the adjacent wire segments. In configuration (a), the magnetization preserves its direction along each wire after traversing the vertex, whereas configuration (b) contains a head-to-head domain wall \cite{hertel_analytic_2015} along one wire and a tail-to-tail wall along another. Configuration (c) is a (double) charged state of the type ``two-in/four-out'' ($-2q$), and configuration (d) contains a quadruple charge structure (``one-in/five-out'', $-4q$). Several equivalent permutations and variations of these configurations are possible, which can be mapped onto each other through rotations, mirror operations, and time-inversion operations $\bm{M}\rightarrow-\bm{M}$. For example, the negatively charged ``one-in/five-out'' structure shown in panel (d) is equivalent to a positively charged configuration of the type ``five-in/one-out'' (not shown). 

In the reference state, all vertices are in the zero-charge magnetic configuration of type I, while the charged state also contains zero-charge configurations of type II alongside double-charge vertex structures. Quadruple-charge vertex structures of the type shown in Fig.~\ref{fig:vertices}d develop only in the random state. These differences in the magnetic configurations result in distinct changes in the high-frequency magnetization dynamics.

\section{High-frequency magnetization dynamics}

\begin{figure}[ht]
\includegraphics[width=\linewidth]{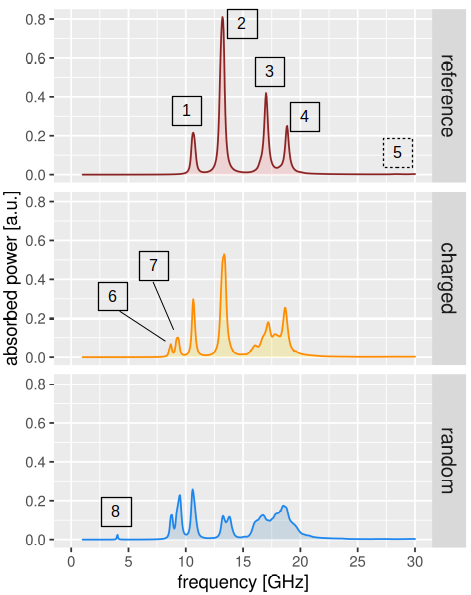}
\caption{\label{fig:spectra}Absorption spectra of the nanowire array for the three magnetic states discussed in the main text. The reference state in the top frame shows sharp resonances at well-defined frequencies. With increasing magnetic disorder, new absorption peaks appear at low frequencies, highlighted in the middle and bottom frame, while the high-frequency peaks near \SI{17}{\giga\hertz} become broader.}  
\end{figure}

Using the numerical methods described in section \ref{sec:method},  we simulate situations in which a low-amplitude harmonic external field is applied to probe the frequency-dependent system's response. At specific frequencies, this field excitation generates resonant oscillations in the magnetic nanowire array. Distinct absorption peaks can be seen in the results displayed in Fig.~\ref{fig:spectra}, which compare the power spectrum of the three zero-field states described above.
The spectra display the frequency-dependence of the power that the system absorbs from the applied rf field~\cite{daquino_micromagnetic_2023}.

\begin{figure*}[ht]
\includegraphics[width=\linewidth]{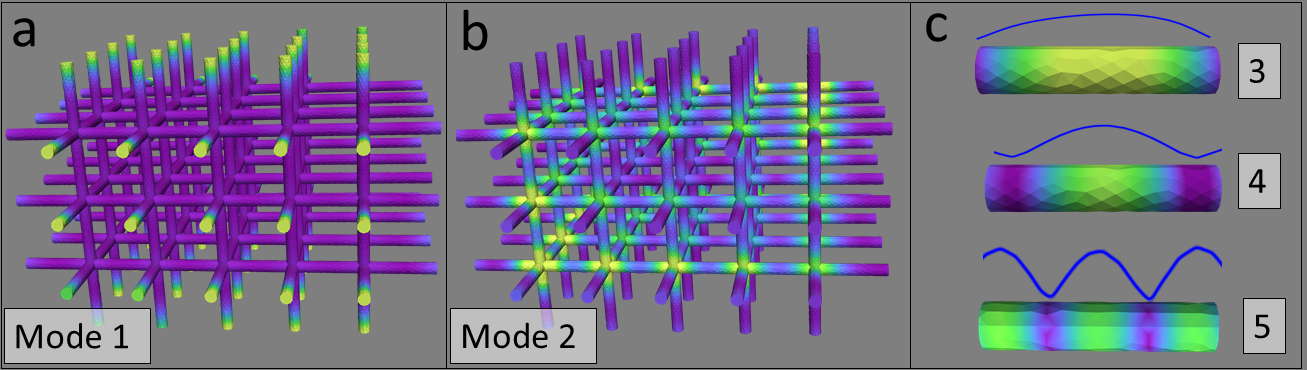}
\caption{\label{fig:reference_modes} Spatial distribution of the oscillation amplitude for the reference state at the frequencies labeled 1-5 in the top frame of Fig.~\ref{fig:spectra}.
The mode profile (a) shows a localization of the oscillations at the dangling, free wire ends at the array surfaces, while frame (b) displays the resonance at the intersection points (vertices). The profiles in panel (c) show different standing-wave modes in individual nanowire array segments. These segments are embedded in the array and are graphically extracted (c) to improve the visualization of the mode profile. The blue lines on top of modes 3, 4, and 5 are numerically determined oscillation amplitude profiles. The color scale indicates the local oscillation strength and ranges from purple (minimum) to yellow (maximum). In each mode, the scale has been adapted to the maximum and minimum amplitude values developing at the given frequency.
}  
\end{figure*}

In the case of the reference state, represented in the frame on the top of Fig.~\ref{fig:spectra}, we observe four well-defined sharp resonances in the absorption spectrum, namely mode \#1 at \SI{10.65}{\giga\hertz}, \#2 at \SI{13.25}{\giga\hertz}, \#3 at \SI{17.0}{\giga\hertz}, and \#4 at \SI{18.85}{\giga\hertz}. A weak fifth absorption peak, whose amplitude is too small to be discerned in the figure, develops at \SI{28.35}{\giga\hertz}.
Each of these resonances can be ascribed to different modes, which develop at specific locations within the array and, in some cases, are characteristic of certain magnetic structures. 
The profiles of these five modes are shown in Fig.~\ref{fig:reference_modes}.
The low-frequency mode \#1 concerns magnetic oscillations at the free ends of the nanowires, i.e., at the outer surface of the array, as shown in Fig.~\ref{fig:reference_modes}a. 
Peak \#2 at the next-higher frequency is due to the resonant excitation of the type-I zero-charge vertices, whose configuration is shown in Fig.~\ref{fig:vertices}. 
The profile of this vertex mode is shown in Fig.~\ref{fig:reference_modes}b. 
In the ordered magnetization state, all vertex structures oscillate at essentially the same frequency, with only minor variations within the array. Upon close inspection, one can identify that the resonance frequency of vertices at specific positions, e.g., at the edges, near the surface, or at the corners,  is slightly different from that of the vertices in the bulk of the array, which we attribute to changes in the local magnetostatic field. 
The remaining modes numbered 3, 4, and 5, develop at higher frequencies and result from standing-wave type oscillations within the nanowires. At these frequencies, the magnetization in the ensemble of nanowire segments oscillates with a specific profile, as shown in Fig.~\ref{fig:reference_modes}c. The mode profiles suggest that the ends of the wires, i.e., the vertex positions, neither act as fixed nor as free boundaries, indicating a frequency dependence of the effective boundary condition \cite{guslienko_effective_2002}.

The resonances discussed so far, with the absorption spectrum shown in Fig.~\ref{fig:spectra}a and the mode profiles displayed in Fig.~\ref{fig:reference_modes}, refer to the magnetically ordered reference state. The spatial distribution of these modes is primarily determined by the geometry, as they appear at specific positions of the three-dimensional structure---the surface, the intersection points, and the nanowire segments. As shown in Fig.~\ref{fig:spectra}b, the spectrum of the charged magnetic state obtained after saturating the sample along the $x$ direction contains further absorption peaks, labeled 6 and 7, in the lower frequency range. These additional peaks are signatures of specific magnetic vertex configurations.

\begin{figure}[ht]
\includegraphics[width=\linewidth]{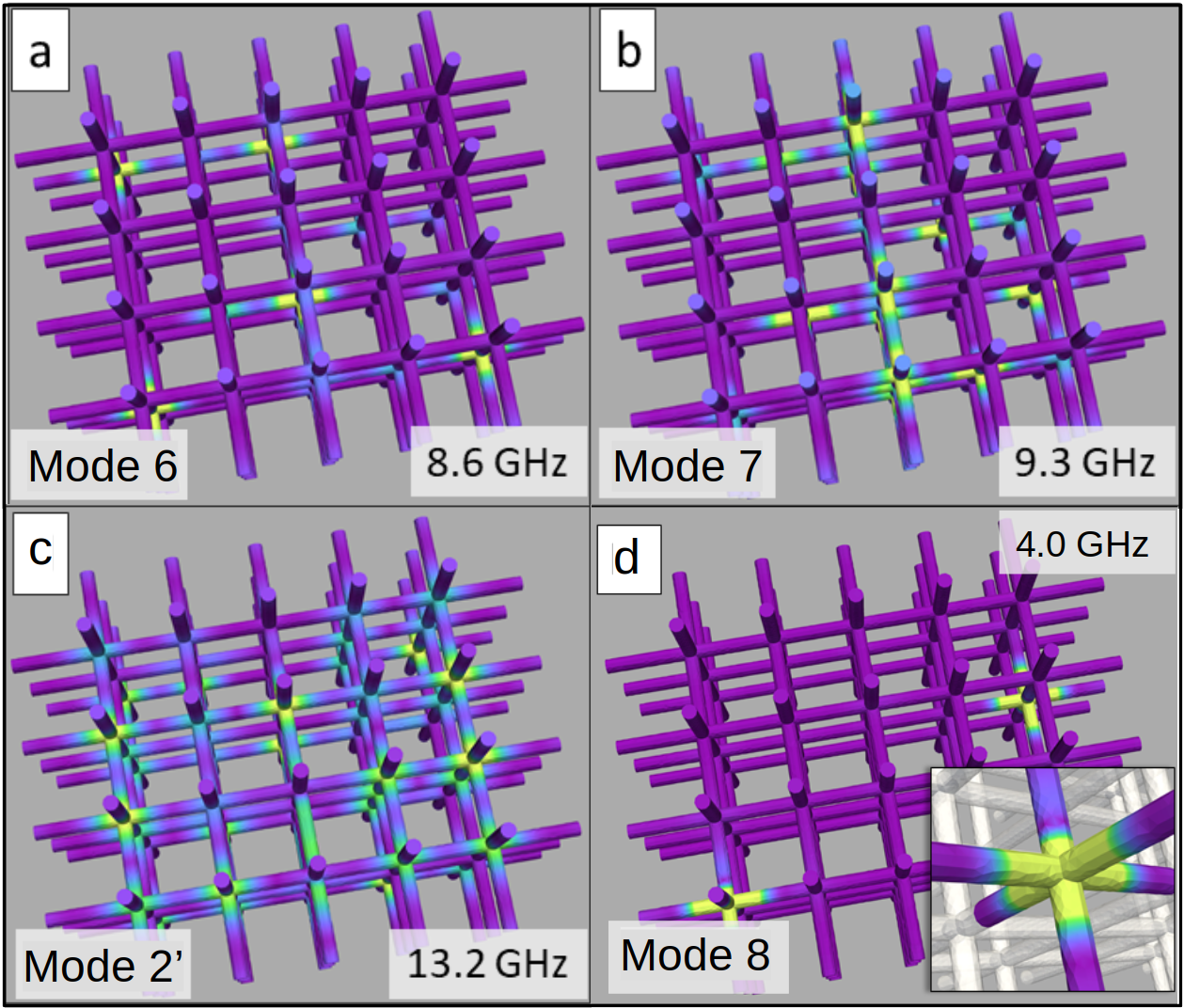}
\caption{\label{fig:vertex_modes} Magnetic modes localized at the vertices. Modes (a), (b) and (c) refer to the charged configuration. Each vertex type becomes resonant at a different frequency. The mode displayed in panel (a) refers to vertices with a configuration as shown in Fig.~\ref{fig:vertices}a, while the localized mode of the vertices with double-charge configuration (cf.~Fig.~\ref{vertex_configs}b) is displayed in panel (b). Panel (c) shows the oscillation of the remaining vertices, which are magnetized in the uncharged configuration of type I and oscillate at the same frequency as in the ordered state, previously labeled as mode 2. 
Panel (d) refers to the disordered (``random'') configuration, which contains two vertices with quadruple charge (cf.~Fig.~\ref{vertex_configs}d) that become resonant at a particularly low frequency. 
}  
\end{figure}

The first peak (mode 6)  refers to the uncharged vertex configuration of type II, shown in Fig.~\ref{fig:vertex_modes}b. Vertices with this magnetic configuration, which combines a head-to-head type transition along one direction and a tail-to-tail one along another, become active at \SI{8.6}{\giga\hertz}, as displayed in Fig.~\ref{fig:vertex_modes}a.
The configuration considered here contains six vertices of this type. At a slightly higher frequency of \SI{9.3}{\giga\hertz}, the charged vertex configurations of the type ``four-in/two-out'' and ``two-in/four-out'', shown in Figs.~\ref{fig:vertices}b, become resonant. In our example, this vertex configuration occurs in a total of 14 vertices, equally distributed in terms of positive and negative charges. Fig.~\ref{fig:vertex_modes}b displays the localized magnetic oscillations of these vertices, constituting mode 7 in the spectrum of Fig.~\ref{fig:spectra}b. 
The previously discussed modes of the reference states also occur in the charged state, such as, e.g., the surface mode of the free ends at \SI{10.65}{\giga\hertz} and the mode of the uncharged type-I vertices at \SI{13.2}{\giga\hertz}. Fig.~\ref{fig:vertex_modes}c shows the spatial distribution of the oscillations of these vertices in the charged configuration. 
The results show that specific absorption peaks can be seen as ``fingerprints'' of different types of vertex configurations, making it possible to deduce the presence of specific magnetic structures by inspecting the absorption spectrum \cite{vanstone_spectral_2022}. To a certain extent, the relative peak heights can moreover serve as an indication for the density of specific configurations in the array, as the intensity of the vertex-related peaks depends on the number of vertices participating in the resonant oscillation. For example, peak 2 of the charged state is diminished compared to the reference state as the latter contains fewer vertices with type-I uncharged configuration. 

The spectrum of the third configuration, the ``random state'' obtained after starting from a randomized initial magnetic strate, is shown in Fig.~\ref{fig:spectra}c. It displays an additional peak labeled 8 at the lower frequency range. This mode is due to the oscillation of vertices with quadruple charge, as shown in Fig.~\ref{fig:vertices}d. Our version of the random state contains two such vertices, one with a positive and the other with a negative charge. The mode profile related to the oscillation of these vertices, which become active at \SI{4.0}{\giga\hertz}, is shown in Fig.~\ref{fig:vertex_modes}d.

In the charged state, and even more so in the random state, the nanowire modes 3 and 4 broaden and merge, evolving into a nearly continuous absorption region between 16 and \SI{19}{\giga\hertz}. This behavior can be explained by magnetostatic fields arising from the charged vertices. Since the vertex charges are sources of magnetostatic fields, they modify the effective field acting along the nanowire segments, which, depending on the charge distribution in the adjacent vertices, may be stronger or weaker than in the ordered state. These variations in the effective field strength modify the nanowire segments' resonance frequency and, thus, lead to a broadening of the absorption peaks.

\section{Conclusion}
We studied the static magnetization and the high-frequency modes developing in a three-dimensional magnetic nanoarchitecture consisting of a regular array of interconnected nanowires through frequency-domain micromagnetic simulations. The system displays distinct absorption peaks at specific frequencies when exposed to a low-amplitude harmonic external magnetic field in the \si{\giga\hertz} frequency range. The simulations reveal that the magnetic oscillations of these resonances are localized at different geometric constituents of the array: the nanowire segments, the intersection points, and the surfaces. In addition to these modes determined by geometric parameters, we find modes that depend on the micromagnetic configuration developing at the vertex points. Such appearance of configuration-dependent vertex modes is similar to analogous effects known from two-dimensional artificial spin-ice systems \cite{gliga_spectral_2013}. The corresponding characteristic absorption lines can thus serve as a means to indirectly detect specific micromagnetic structures \cite{vanstone_spectral_2022}, with the peak height indicating their density within the array. By combining three-dimensional aspects of magnonic crystals and artificial spin-ice systems, regular arrays of magnetic nanowires of the type studied in this article provide a variety of micromagnetic properties, particularly regarding the high-frequency magnetization dynamics, which could render these artificial material types interesting for reprogrammable magnonic applications \cite{gliga_dynamics_2020}.

\begin{acknowledgments}

This work was funded by the French National Research Agency (ANR) through the Programme d'Investissement d'Avenir under contract ANR-11-LABX-0058\_NIE and ANR-17-EURE-0024 within the Investissement d’Avenir program ANR-10-IDEX-0002-02. The
authors acknowledge the High Performance Computing center of
the University of Strasbourg for supporting this work by providing access to computing resources.

\end{acknowledgments}


\end{document}